\documentclass[12pt]{article}
\pdfoutput=1
\usepackage[utf8]{inputenc}
\usepackage{amsmath}
\usepackage{amssymb}
\usepackage{units}
\usepackage{graphicx}
\usepackage{slashed}
\usepackage{array}
\usepackage{xcolor}
\usepackage[leftcaption,raggedright]{sidecap}
\usepackage{caption}
\usepackage{subfigure,dcolumn}
\usepackage[hidelinks]{hyperref}

\def\Title#1{\begin{center} {\Large #1 } \end{center}}
\def\Author#1{\begin{center}{ \sc #1} \end{center}}
\def\Address#1{\begin{center}{ \it #1} \end{center}}

\newenvironment{Abstract}{\begin{quotation}  }{\end{quotation}}
\newenvironment{Presented}{\begin{quotation} \begin{center} 
             PRESENTED AT\end{center}\bigskip 
      \begin{center}\begin{large}}{\end{large}\end{center} \end{quotation}}

\def\beq{\begin{equation}}
\def\eeq#1{\label{#1}\end{equation}}
\def\eeqn{\end{equation}}

\def\beqa{\begin{eqnarray}}
\def\eeqa#1{\label{#1}\end{eqnarray}}
\def\eeqan{\end{eqnarray}}

\let\bar=\overbar

\def\Dslash{\not{\hbox{\kern-4pt $D$}}}
\def\dslash{\not{\hbox{\kern-2pt $\del$}}}

\def\msb{{\bar{\ssstyle M \kern -1pt S}}}

\begin{document}
\begin{titlepage}

\vfill
\Title{Design and Integration of JUNO-OSIRIS}
\vfill
\Author{Narongkiat Rodphai and Zhimin Wang \\ on behalf of JUNO Collaboration}
\Address{Institute of High Energy Physics, Beijing, 100049, China \\ University of Chinese Academy of Sciences, Beijing, 100049, China
}
\vfill
\begin{Abstract}
The Jiangmen Underground Neutrino Observatory (JUNO) is a neutrino detection experiment characterized by an acrylic sphere, measuring 35.4 meters in diameter, containing 20,000 tons of liquid scintillator. This sphere is encompassed by as many as 17,600 photomultiplier tubes (PMTs) with a 20-inch diameter, achieving an overall coverage of 77.9\%. With these impressive capabilities, the JUNO experiment aims to achieve an exceptional effective energy resolution of 3\% at 1 MeV.
\par The Online Scintillator Internal Radioactivity Investigation System of JUNO (OSIRIS) serves as a precursor detector, around 100 meters aside in horizontal of the site of the JUNO detector underground, tasked with monitoring and examining the purity of the liquid scintillator prior to its transfer to the JUNO central detector. The OSIRIS, also as a pre-detector of JUNO, is constructed with a cylindrical acrylic vessel designed to hold 18 tons of liquid scintillator. It is situated within a 9-meter height cylindrical tank filled with 550 tons of pure water. This detector was specifically engineered to search for the fast coincidence decays of $^{214}$Bi -$^{214}$Po and $^{212}$Bi -$^{212}$Po in the decay chains of $^{238}$U and $^{232}$Th, respectively. 
OSIRIS has been designed to reach a sensitivity of 10$^{-16}$ g/g for U/Th to test scintillator radiopurity to the level required for the detection of solar neutrinos. Additionally, 64 20-inch microchannel plate PMTs (MCP PMTs) were positioned around the liquid scintillator vessel to observe incoming interactions, along with an additional 12 20-inch MCP PMTs for the water Cherenkov muon veto system.
\par The OSIRIS pre-detector has been fully constructed and integrated,  providing a plenty of preliminary results from the air runs. The next step in the process involves commencing the filling soon, aiming to involve an in-depth examination of the impurities within the liquid scintillator.

\end{Abstract}
\vfill
\begin{Presented}
NuPhys2023, Prospects in Neutrino Physics\\
King's College, London, UK,\\ December 18--20, 2023
\end{Presented}
\vfill
\end{titlepage}
\def\thefootnote{\fnsymbol{footnote}}
\setcounter{footnote}{0}

\section{JUNO Experiment Design}
\begin{figure*}[!ht]
	\centering
	\subfigure[]{
        \includegraphics[width=0.45\hsize]{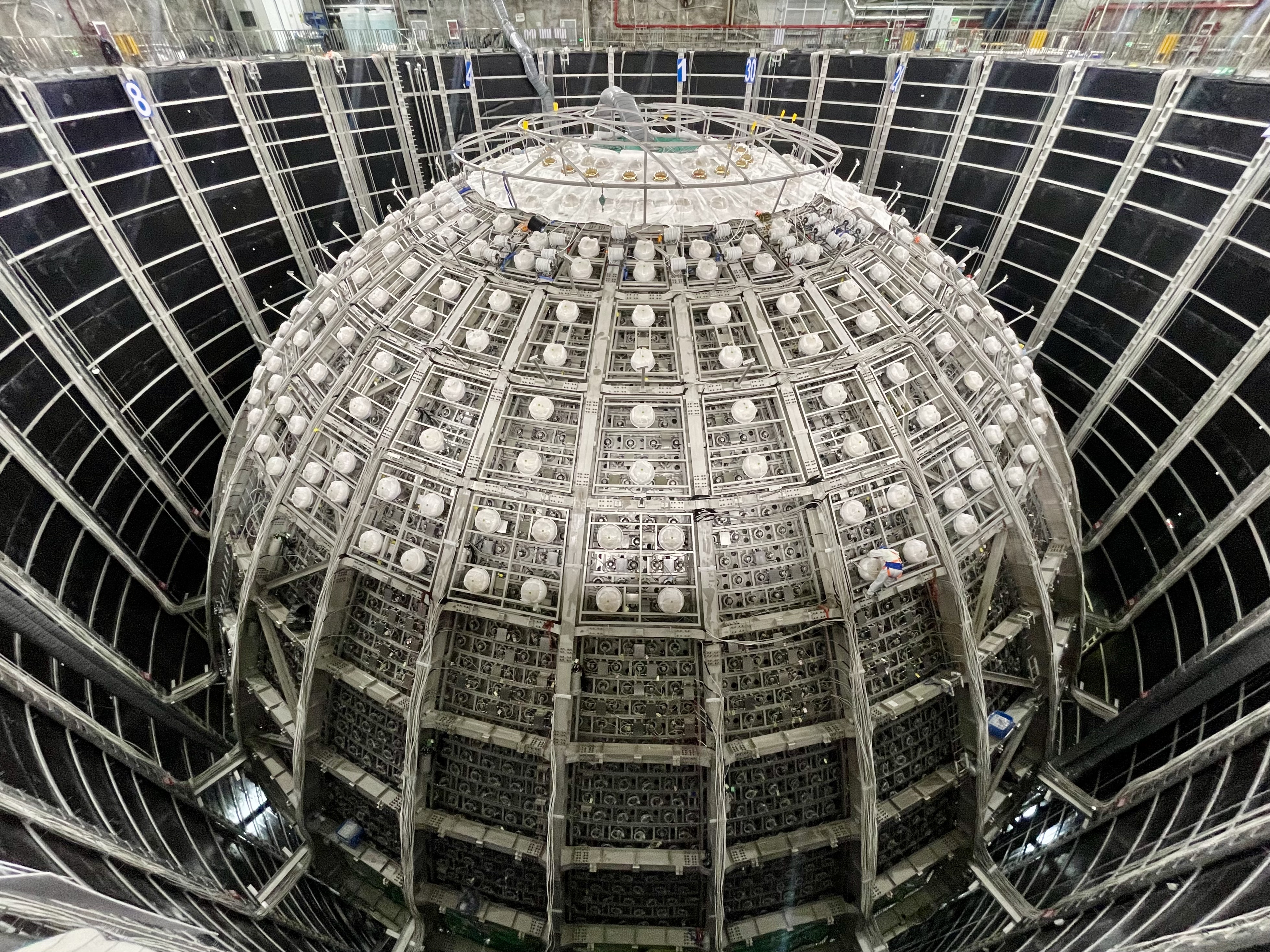}
        \label{fig:juno1}
    }
    \quad
	\subfigure[]{
        \includegraphics[width=0.45\hsize]{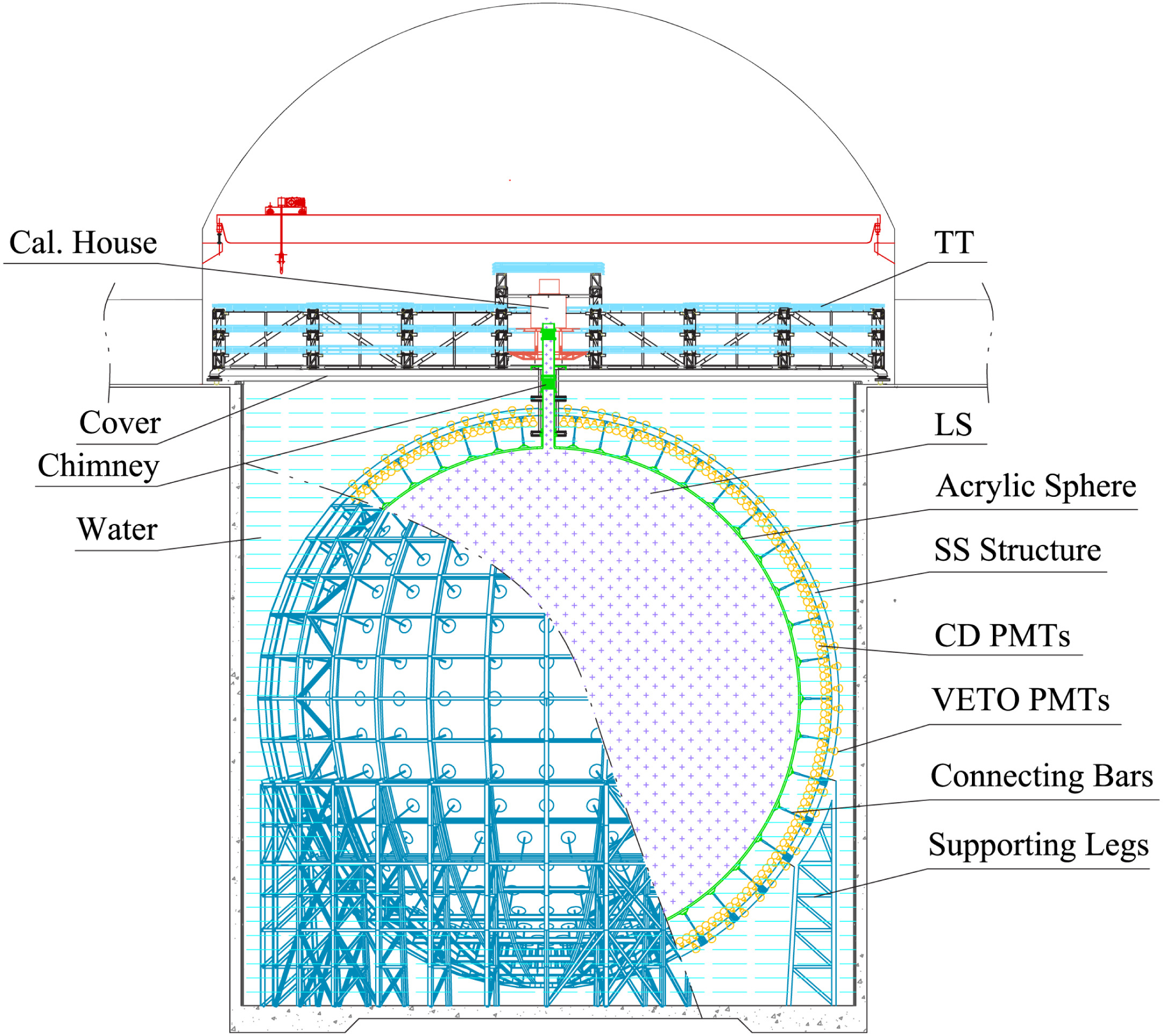}
        \label{fig:juno2}
    }
    \caption{The JUNO central detector inside the cylindrical water pool under construction (a) and the overall scheme of the JUNO detector (b)}
    \label{fig:juno}
\end{figure*}

   The Jiangmen Underground Neutrino Observatory (JUNO) is now under construction underground with 700 meters overburden of rock in Jiangmen, Guangdong, the southern China, aiming to observe incoming neutrinos from several sources. The main objective is to determine the neutrino mass hierarchy and neutrino oscillation parameters, and also other neutrino physics problems \cite{1}. 
   The central detector (CD) was constructed as an enormous acrylic sphere with 35.4 meters in diameter, containing with over 20,000 tons of liquid scintillator (LS). This CD will be constructed and submerged in a cylindrical water pool with 35,000 ton of high purity water, as the VETO detector as shown in Figure \ref{fig:juno}. There will be over 17,600 20-inch photomultiplier tubes (PMTs) and 25,600 3-inch PMTs, installed all over the CD with an overall coverage of 77.9\% to observe an interaction between incoming neutrinos and liquid medium according to the inverse beta decay (IBD) and another 2,400 20-inch PMTs used for the water Cherenkov VETO detector. With these impressive capabilities, the 20-inch PMTs \cite{2} and liquid scintillator both play a crucial role for the JUNO experiment achieve an exceptional effective energy resolution of 3\% at 1 MeV.

   \par
\begin{figure}[!hbt]
    \centering
    \includegraphics[width=0.85\hsize]{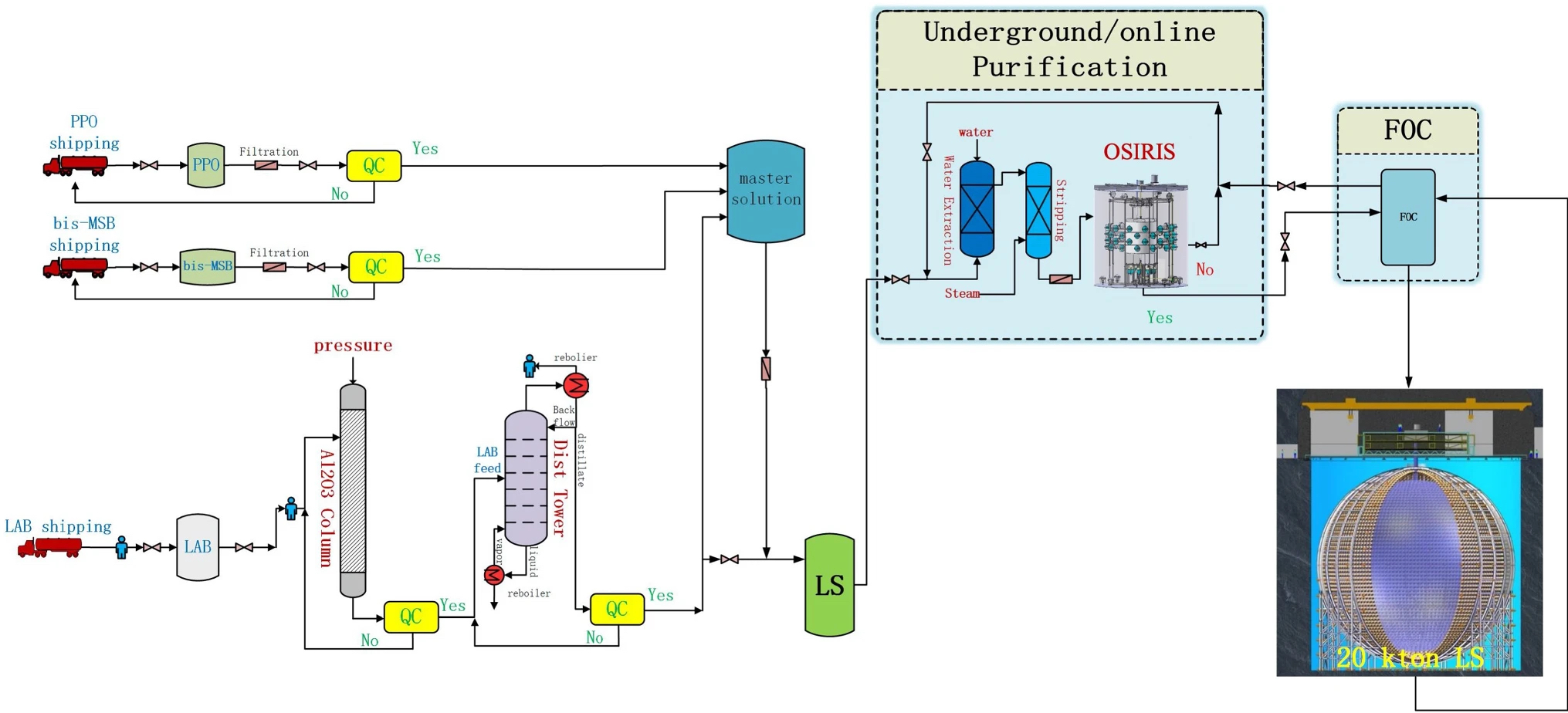}
    \caption{Flowchart of the liquid scintillator processing system\cite{3}}
    \label{fig:LS}
\end{figure}

 20,000 tons of LS, used in this experiment, contains three components: Linear Alkyl Benzene (LAB), 2,5-diphenyloxazole (PPO), and 1,4-bis (2-methylstyryl) benzene (bis-MSB), which requires the U/Th purity of 1 $\times$ 10$^{-15}$ g/g for the reactor neutrino studies and 1 $\times$ 10$^{-17}$ g/g for the solar neutrino studies. These liquid will be prepared by the LS processing system after several stages distributed on the ground as illustrated in Figure \ref{fig:LS}, and transferred to underground for testing the contamination with the water extraction and gas stripping systems before transferring to the Online Scintillator Internal Radioactivity Investigation System (OSIRIS) detector \cite{3}.
   
\section{OSIRIS Detector}
\begin{figure*}[!ht]
	\centering
	\subfigure[]{
        \includegraphics[width=0.45\hsize]{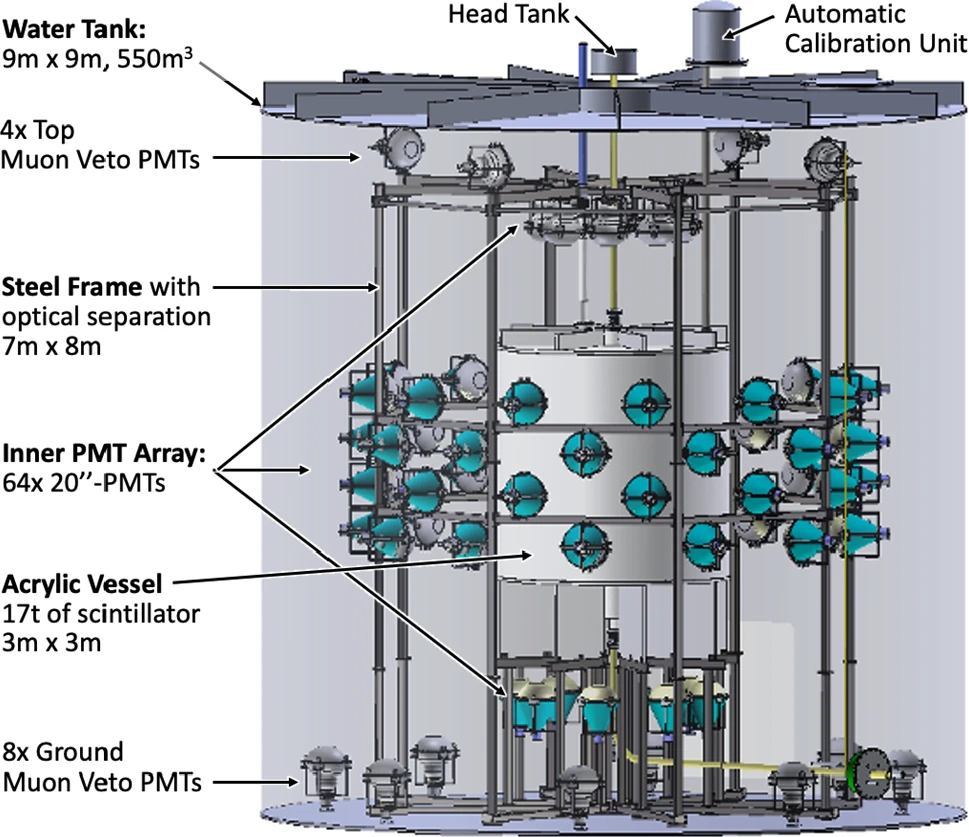}
        \label{fig:osiris1}
    }
    \quad
	\subfigure[]{
        \includegraphics[width=0.45\hsize]{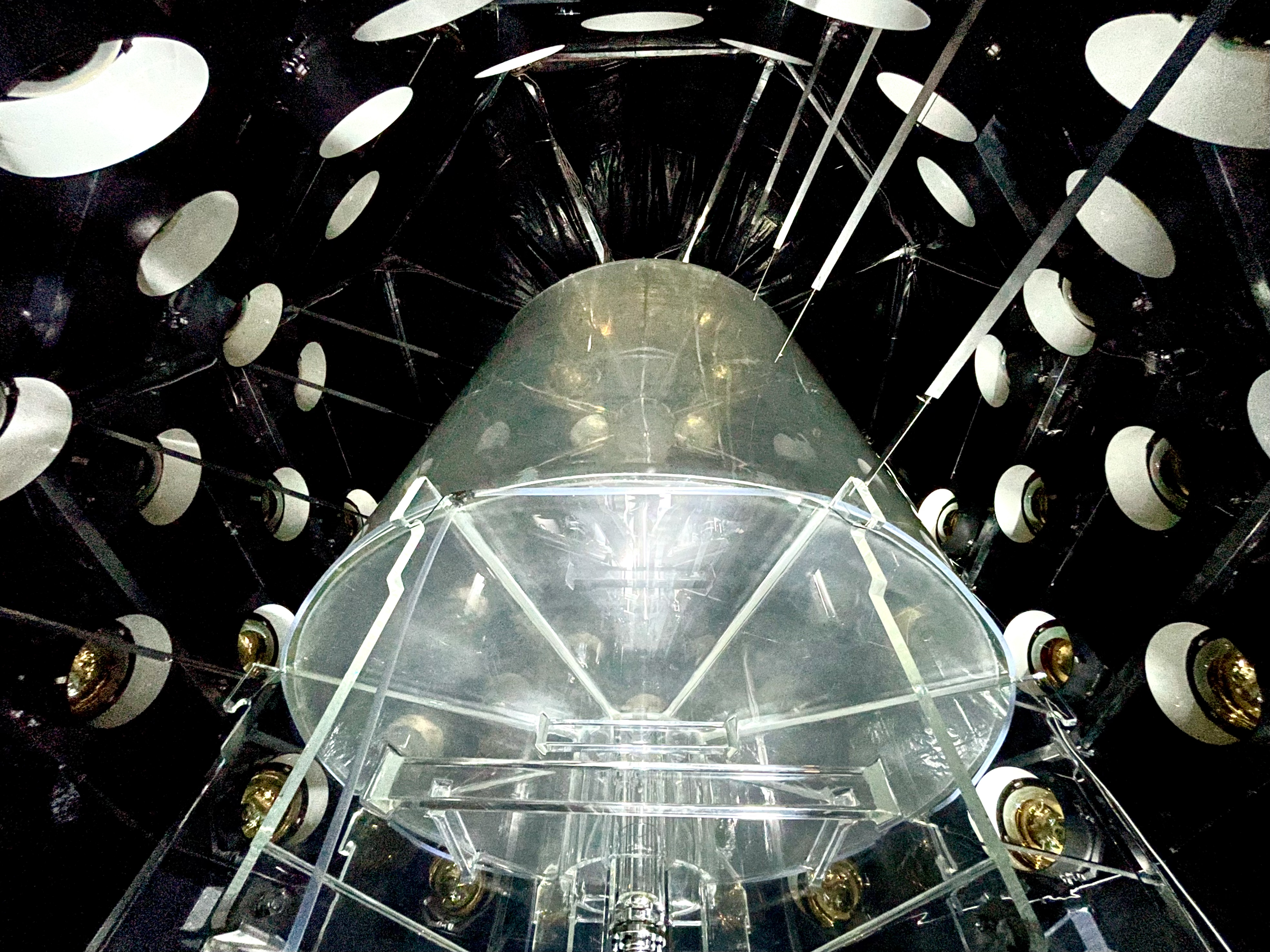}
        \label{fig:osiris2}
    }
    \caption{The overall OSIRIS detector scheme \cite{2} (a) and an inside of OSIRIS detector with cylindrical acrylic vessel surrounded by 20-inch PMTs (b)}
    \label{fig:osiris}
\end{figure*}

 The Online Scintillator Internal Radioactivity Investigation System (OSIRIS) serves as a precursor detector, tasked with monitoring and examining the LS purity prior to its transfer to the JUNO Fluid Connection and Control System (FOC) and CD. The OSIRIS detector is constructed with a 3-meter diameter cylindrical acrylic vessel designed to hold 18 tons of LS as shown in Figure \ref{fig:osiris}. It is situated within a 9-meter height cylindrical tank filled with 550 tons of pure water. This detector was specifically engineered to search for the fast coincidence decays of $^{214}$Bi -$^{214}$Po and $^{212}$Bi -$^{212}$Po in the decay chains of $^{238}$U and $^{232}$Th, respectively. The projected sensitivity is 10$^{-16}$ g/g, sufficient to test radiopurity levels for reactor (and even solar) neutrino detection \cite{3}. Additionally, 64 20-inch microchannel plate (MCP) PMTs were positioned around the LS vessel to observe interactions in LS, along with an additional 12 MCP PMTs for the water Cherenkov muon veto system. 
 All the PMTs are equipped with cones offering magnetic shielding and a front cylinder reflector as shown in Figure \ref{fig:osiris2}, and equipped with JUNO 1F3 electronics \cite{4,5}.
 
\section{OSIRIS Water Filling Process}
\begin{figure*}[!ht]
	\centering
	\subfigure[]{
        \includegraphics[height=0.32\hsize]{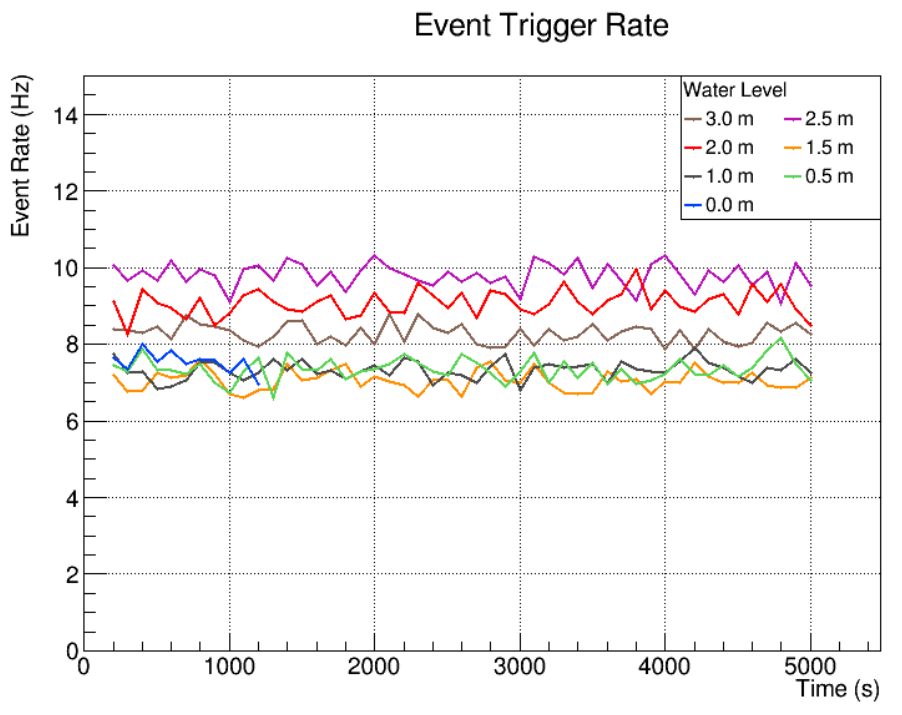}
        \label{fig:result1}
    }
    \quad
	\subfigure[]{
        \includegraphics[height=0.32\hsize]{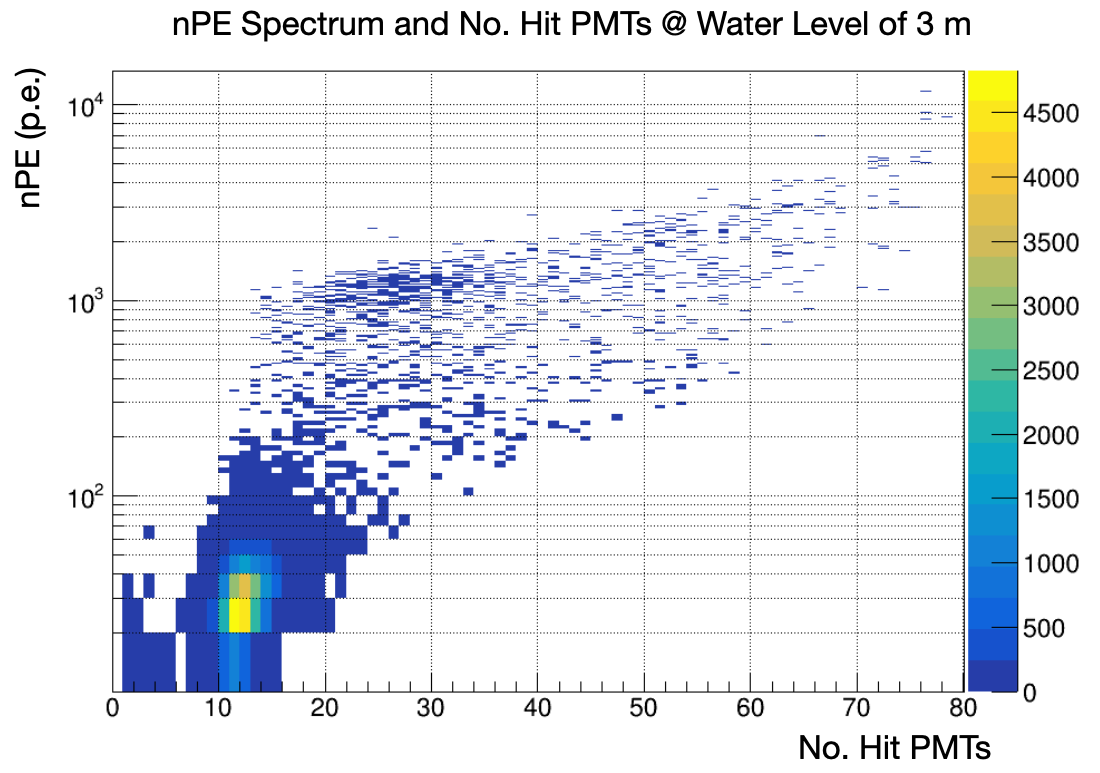}
        \label{fig:result2}
    }
    \caption{The preliminary results of the event trigger rate with different water level (a) and the nPE spectrum with different No. of Hit PMTs (b)
}
    \label{fig:result}
\end{figure*}

Late of November 2023, we started to fill the detector with the first pure water batch of 185 tons, reaching about 3 meters, just lower than the bottom of the acrylic vessel or one third of the detector which 16 PMTs were under water. We run the data taking with a threshold of 11 PMTs fired in 64 ns out of 76 PMTs in total, to observe the event waveform (also another triggerless data stream only with (Start time, Charge) information) during this water filling process and observed the event rate of 7-10 Hz as shown in Figure \ref{fig:result1}. 
Increasing the water level also increased the probability from  the surrounded background signals, interacting with the water inside.
We also applied the number of Hit PMTs as threshold to determine the nPE spectrum correlation (Figure \ref{fig:result2}). 
This shows us the most nPE spectrum region were detected by $\sim$12 Hit PMTs, which is our multiplicity threshold and mostly generated by the general events and radioactivity background.
However, 
we detect a population of events with well over 100 photoelectrons (p.e.) in total per event that are associated to the Cherenkov signal of cosmic muons.
It is calculated the classified muon event rate of about 0.3 Hz, which is basically consistent with the pre-estimated muon rate of $\sim$10.8 per m$^2$ per hour \cite{1}.   

\section{Conclusions and Outlook}
The ongoing OSIRIS water filling process is aimed at filling the entire detector to facilitate the acquisition of essential information and characteristics of both event and background signals. This process involves the infusion of pure water and liquid scintillator, PMT and electroncis, rendering the OSIRIS detector a precursor detector for the JUNO experiment. 
Beyond its primary purpose to investigate the radiopurity of the liquid scintillator, the OSIRIS detector serves as a test bed for detector subsystems to be used in JUNO later on and for the study of technologies for future upgrades.

\providecommand{\href}[2]{#2}\begingroup\raggedright\endgroup

\end{document}